\documentclass[11pt]{article}
\usepackage{epsfig,cite}
\usepackage{latexsym}
\usepackage{threeparttable}

\newcommand{\beq}{\begin{eqnarray}}
\newcommand{\eeq}{\end{eqnarray}}
\newcommand{\be}{\begin{eqnarray*}}
\newcommand{\ee}{\end{eqnarray*}}

\newcommand{\D}{{\cal D}}
\newcommand{\Pom}{{\hspace{ -0.1em}I\hspace{-0.25em}P}}
\newcommand{\Reg}{{\hspace{ -0.1em}I\hspace{-0.25em}R}}
\newcommand{\vp}{{\gamma^*}}
\newcommand{\qqbar}{q \overline{q}}

\newcommand{\xP}{x_\Pom}
\newcommand{\FD}{F_{2 \D}^{(3)} }

\begin{document}
\title{A unitarized model of inclusive and diffractive DIS with $Q^2$-evolution}
\author{
\begin{tabular}{c}
N\'estor~Armesto,$^1$ \hspace{1.em} Alexei~ B.~Kaidalov,$^2$ \\ Carlos~A.~Salgado,$^1$ \hspace{0.5em} and \hspace{0.5em} Konrad~Tywoniuk$^1$ \\
\\
$^1$ Departamento de F\'isica de Part\'iculas and IGFAE, \\
Universidade de Santiago de Compostela, \\ 
15706 Santiago de Compostela, Galicia, Spain \\ \\
$^2$ Institute of Theoretical and Experimental Physics,\\
117259 Moscow, Russia
\end{tabular}
}
\date{\today}
\maketitle

\begin{abstract}
We discuss the interplay of low-$x$ physics and QCD scaling violations by extending the unified approach describing inclusive structure functions and diffractive production in $\vp p$ interactions proposed in previous papers, to large values of $Q^2$. We describe the procedure of extracting, from the non-perturbative model, initial conditions for the QCD evolution that respect unitarity. Assuming Regge factorization of the diffractive structure function, a similar procedure is proposed for the calculation of hard diffraction. The results are in good agreement with experimental data on the proton structure function $F_2$ and the most recent data on the reduced diffractive cross section, $\xP \sigma_r^{\D(3)}$. Predictions for both $F_2$ and $F_L$ are presented in a wide kinematical range and compared to calculations within high-energy QCD.
\end{abstract}

\section{Introduction}

The interaction of a highly energetic virtual photon with a nucleon, or a nucleus, probes the high-energy limit of strong interactions. In this situation it is important both to find the correct degrees of freedom of the interaction and to preserve unitarity of the cross section. In the infinite momentum frame the latter comes about, in the limit of large gluon density, as non-linear gluon interactions which tame the growth of parton distributions and eventually leads to saturation. In the target rest frame the same phenomenon is manifested in terms of multiple scattering. 

Experimental data indicate a fast growth of the $\vp p$ cross section accompanied by a large diffractive cross section at  the highest available energies \cite{Adams:1996gu,Arneodo:1996qe,Abt:1993cb,Ahmed:1995fd,Aid:1996au,Adloff:1997mf,Adloff:1999ah,Adloff:2000qk,Derrick:1993fta,Derrick:1994sz,Derrick:1995ef,Derrick:1996hn,Breitweg:1997hz,Breitweg:1998dz,Breitweg:2000yn,Chekanov:2001qu,Aaron:2008tx,Chekanov:2009na,Chekanov:2008fh,Adloff:1997mi,Aktas:2006hy}. This calls for a unified treatment of several effects: a proper unitarization of the total cross section at low momentum scales in  conjunction with a correct treatment of QCD scaling violations. While present day data give a rich insight into the latter, the question of whether saturation effects have been observed is still under debate. All the same, these effects are of crucial importance when extrapolating to higher energies, deep inelastic scattering (DIS) off nuclei and heavy-ion collisions.

DIS measurements are typically used to constrain universal parton distribution functions (PDFs) through an analysis of QCD scaling violations. The most recent state-of-the-art fits based on the DGLAP evolution equations up to NNLO have been presented in \cite{Martin:2009iq,Nadolsky:2008zw}. Regrettably, one has no perturbative theoretical control of the initial conditions and, thus, extrapolations to kinematical regions yet unexplored by experiments are affected by significant uncertainties, e.g. as obtained by recent PDF fits using neural networks \cite{Ball:2008by}.

An established approach to the problem of high-energy scattering in QCD is the dipole model \cite{Mueller:1989st,Nikolaev:1990ja}. One can show that at very high energies the $\vp p$ scattering can be modeled as a convolution of the wave function of a virtual photon fluctuating into a $q\bar{q}$ dipole and the subsequent dipole-proton cross section. While the former is known exactly from QED, the latter can be treated theoretically (approximated e.g. by two-gluon exchange) or simply parameterized. Still, the dipole model contains several weaknesses which limit its range of applicability, such as lack of scaling violations owing to QCD evolution. Furthermore, the model is only valid for small dipole sizes and at low-$x$. 

Using perturbative arguments one can show that the strength of the $\qqbar-N$ cross section should scale with the transverse area of the dipole \cite{Mueller:1989st,Nikolaev:1990ja}. A particular simple parameterization of this cross section was suggested in \cite{GolecBiernat:1998js}. The so-called GBW dipole cross section encodes the characteristic features of parton saturation by assuming an $x$-dependent damping of large-size dipoles, or, in other words, a saturation scale $Q^2_s \propto x^{-\Delta}$. A fit to experimental data resulted in a value $\Delta \approx 0.28$. The model was extended to describe diffraction in \cite{GolecBiernat:1999qd}.

In fact, the effective energy dependence of the cross section, $\Delta_{eff} = d  \ln F_2 \big/ d \ln (1/x)$, changes from a quite small value at low-$Q^2$, consistent with the so-called soft pomeron, to a steady growth in the perturbative region. The dipole model incorporates some scaling violations in the $\vp-\qqbar$ wave function which mimic the QCD behavior, but fails to match with the DGLAP equations at high-$Q^2$ data due to the lack of scale dependence of the $\qqbar-N$ cross section. The authors in \cite{Bartels:2002cj} attempt to include these by making use of the connection between the dipole cross section and the gluon distribution function at leading logarithmic accuracy. At the initial scale they include solely a non-zero gluon PDF and perform a LO QCD fit to constrain its parameters. An impact parameter dependence was also introduced in a later extension of the model \cite{Kowalski:2003hm}. Calculations of diffraction within the improved dipole model have also been presented \cite{Marquet:2007nf}.

Recently, a general evolution equation for high-energy QCD has been derived using renormalization group techniques, the so-called JIMWLK equation  \cite{JalilianMarian:1997gr,JalilianMarian:1997dw,Kovner:2000pt,Weigert:2000gi,Iancu:2000hn,Ferreiro:2001qy,Balitsky:1995ub}. It can be approximated, with great accuracy, by the significantly less complicated BK equation \cite{Balitsky:1995ub,Kovchegov:1999yj}. Unfortunately, calculations at leading order result, for realistic values of the strong coupling constant, in an $x$-dependence of the saturation scale significantly larger than the $\Delta$ extracted phenomenologically \cite{GolecBiernat:1998js}. Sub-leading effects, such as energy-momentum conservation and inclusion of running of the strong coupling constant, are believed to tame this growth \cite{Albacete:2004gw,Chachamis:2004ab}. It was not until recently that a complete numerical solution \cite{Albacete:2007yr} of the BK equation with running coupling was presented \cite{Balitsky:2006wa,Kovchegov:2006vj,Gardi:2006rp}. In \cite{Albacete:2009fh} a set of parameters related to the initial distribution was fixed to obtain a good agreement with $F_2$ and $F_L$ at $x < 0.01$ and predictions down to $x = 10^{-12}$ were made for a wide range of $Q^2$.

In spite of the successes, there are many aspects of the theoretical computations which are still to be fully understood, such as the impact parameter dependence and the description of diffraction. In the absence of a unified QCD approach to the entirety of $\vp N$ ($\vp A$) processes, a useful guidance for investigating the connection between non-perturbative and perturbative aspects of DIS valid for extrapolations to extremely small momentum fractions, $x$, can be found in terms of the Reggeon calculus \cite{Gribov:1968fc} with a supercritical pomeron, $\Delta_\Pom = \alpha_\Pom - 1 > 0$, and the partonic picture of $\vp N$ interactions. Multi-reggeon exchanges are included so as to satisfy $s$-channel unitarity. In a particular realization of these models, in the $\vp$ wave-function, one distinguishes explicitly between a large ($L$) and a small ($S$) component \cite{Capella:2000pe,Capella:2000hq}. The former interacts strongly even at high-$Q^2$ but is quite rare, while the latter interacts according to $r\sim 1/Q$. Thus both components have a leading $1/Q^2$ dependence of the total cross section while the $L$-component gives the leading contribution to diffraction ($1/Q^2$ vs. $1/Q^4$). Large-mass diffraction is included through triple-reggeon interactions. In this work we will follow the approach of \cite{Capella:2000hq}, where the small component was cast in the form of the dipole model. The model gave a simultaneous description of inclusive $F_2$ and diffraction in the region of $0 < Q^2 \leq 5-10$ GeV$^2$, and was used to predict the structure functions at very low $x$. In \cite{Armesto:2003fi} these results were used to predict nuclear shadowing calculated in the Glauber-Gribov theory, in good agreement with data. In Sec.~\ref{sec:CFSKInclusive} we give a detailed description of the model (formulae for diffraction are given in Appendix~\ref{app:CFSKdiffractive}). 

With the advent of high-energy colliders, such as HERA and LHC, and the planned electron-hadron experiments \cite{eic,lhec2}, the need for low-$x$ structure functions for nucleons and nuclei at high-$Q^2$ have arisen. This motivates an extension of the model mentioned above \cite{Capella:2000hq} to the perturbative regime by the inclusion of QCD scaling violations. We describe a prescription for extracting the initial conditions at leading order for the DGLAP equations from the non-perturbative model both for inclusive $F_2$ and diffraction. In the former case, this procedure does not involve new parameters. The situation for the inclusive diffractive cross section is more complex, because it involves both more complicated reggeon exchanges and additional variables in the problem. For the proper description of data in the whole $\beta$ and $\xP$ region we identify explicitly pomeron and reggeon contributions to diffraction. One can then invoke a supplementary factorization of variables, the so-called Regge factorization \cite{Ingelman:1984ns}, which allows for a comprehensible QCD analysis. In the reggeon case, we identify and include missing diagrams which are crucial for a proper description of data. Details on the initial conditions and subsequent QCD evolution in the inclusive and diffractive cases are given in Secs.~\ref{sec:QCDinclusive} and \ref{sec:QCDdiffractive}, respectively. 

Thus, equipped with properly unitarized initial conditions for the DGLAP evolution equations we obtain leading-order structure functions and PDFs for the proton down to $x \sim 10^{-8}$ at high-$Q^2$. The resulting $F_2$ and $\xP F_{2 \D}^{(3)}$ are shown to be in good agreement with the most recent experimental data. We also compute the longitudinal structure function within the dipole model using the perturbative gluon PDF thus obtained in Sec.~\ref{sec:FL}. Comparisons are made with the recently computed solution of the running-coupling BK equation \cite{Albacete:2009fh}. Finally, we present our conclusions in Sec.~\ref{sec:Conclusions}.

\section{Brief description of the CFSK model \label{sec:CFSKInclusive}}
At small-$x$, the total cross section for $\vp p$ interactions is related to the structure function $F_2(x,Q^2)$ by
\beq
\label{eq:TotalCrossSection}
\sigma_{\vp p}^{(tot)} (W^2,Q^2) \,=\, \frac{4 \pi^2 \alpha_{e.m.}}{Q^2} \, F_2(x,Q^2) \;,
\eeq
where $x = Q^2/(W^2 + Q^2)$, $W = \sqrt s$ is the invariant mass of the produced hadronic system and $Q^2$ is the virtuality of the photon. In the model of  \cite{Capella:2000hq}, denoted in the following as CFSK, the total cross section was written as a sum of two contributions 
\beq
\sigma_{\vp p}^{(tot)} (s,Q^2) &=& \int d^2b \, \sigma_{\vp p}^{(tot)}(b,s,Q^2) \;, \\
\label{eq:CrossSectionDecomposition}
\sigma_{\vp p}^{(tot)}(b,s,Q^2) &=& g^2_L(Q^2) \sigma_L^{(tot)}(b,s,Q^2) \,+\, \sigma_S^{(tot)}(b,s,Q^2) \;.
\eeq
The first term on the right hand side of eq.~(\ref{eq:CrossSectionDecomposition}) describes the non-perturbative interaction of large-size partonic configurations of the virtual photon with the target, while the second term describes the interaction of small-size configurations. The function $g^2_L(Q^2)$, determining the coupling of the $\vp$ to the large $\qqbar$ dipole, was chosen in the form
\beq
g^2_L(Q^2) \,=\, \frac{g^2_L(0)}{1 + Q^2/m^2_L} \;,
\eeq
vanishing at high $Q^2$ while the $Q^2$-dependence of the small-size component is an inherent characteristic of the dipole model.

\begin{figure}[t!]
\begin{center}
\includegraphics[width=0.45\textwidth]{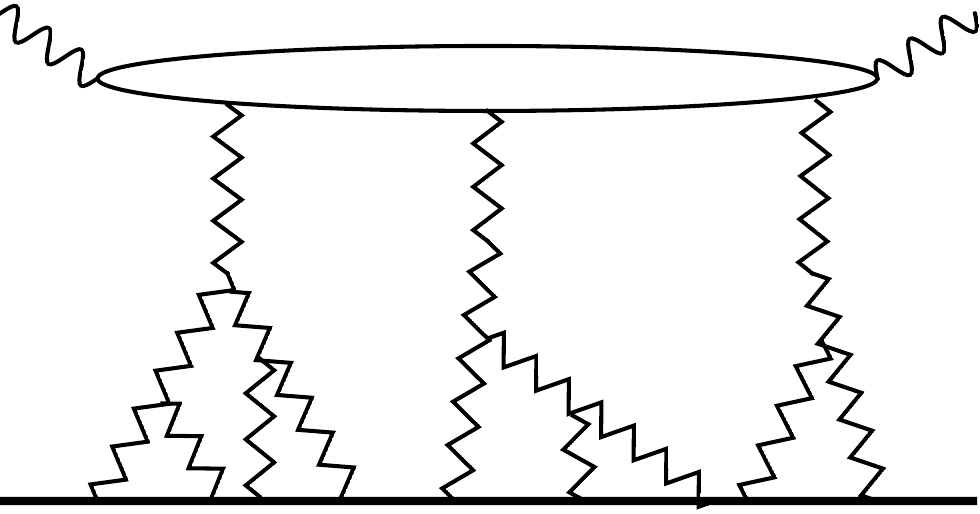}
\caption{A generic reggeon diagram included in the CFSK model.}
\label{fig:ReggeonDiagram}
\end{center}
\end{figure}
The cross section of the $L$-component in impact parameter space is cast in the quasi-eikonal form (to include proton dissociation)
\beq
\centering
\label{eq:LargeComponent}
\sigma_L^{(tot)}(b,s,Q^2) \,&=&\, \frac{1 - \exp \left( - C \chi_L(b,s,Q^2) \right)}{2 C} \;,
\eeq
where the function $\chi_L$ accounts for multiple pomeron ($\Pom$) and reggeon ($\Reg =$ \{$f$, $A_2$,... \}) exchanges and triple-reggeon interactions, as follows
\beq
\label{eq:TotalChi}
\chi_L(s,b,Q^2) \,&=&\, \frac{\chi_{L 0}^\Pom (b,\xi)}{1 + a \chi_3(s,b,Q^2)} \,+\, \chi_{L 0}^\Reg (b,\xi) \;.
\eeq
The first term in eq.~(\ref{eq:TotalChi}) corresponds to a summation of fan-type diagrams, also called the Schwimmer model \cite{Schwimmer:1975}. The constant $a$ is defined as $a = g_{pp}^\Pom(0) r_{\Pom \Pom \Pom} / 16 \pi$, where $g_{pp}^\Pom(0)$ is the proton-pomeron coupling and $r_{\Pom \Pom \Pom}$ is the triple-pomeron coupling. The eikonal functions $\chi_{L 0}^k$ ($k = \Pom, \Reg$) are written in the standard Regge form
\beq
\chi_{L 0}^k (b,\xi) \,=\, \frac{C_L^k}{\lambda_{0 k}^L(\xi)} \, \exp \left( \Delta_k \xi \,-\, \frac{b^2}{4 \lambda_{0 k}^L(\xi)} \right) \;,
\eeq
where
\beq
\label{eq:PomParameters}
\Delta_k = \alpha_k(0) -1 \;\;, \;\; \xi = \ln \frac{s+Q^2}{s_0 + Q^2} \;\; , \;\; \lambda^L_{0 k} = R^2_{0 k  L} + \alpha'_k \xi
\eeq
and $\alpha_k(0)$, $\alpha'_k$ are the intercept and slope of the corresponding trajectories, respectively. The function $\xi$ is chosen such that it behaves as $\ln (1\big/x) $ at high-$Q^2$ and $\ln (s\big/ s_0)$ at $Q^2=0$. The triple-reggeon interaction term, $\chi_3(s,b,Q^2)$ is 
\beq
\label{eq:chi3}
\chi_3 (s,b,Q^2,\beta) &=& \frac{1}{\lambda_\Pom \left(\xP \right)} \, \exp \left[ - \frac{ b^2}{4 \lambda_\Pom \left( \xP \right)} \right] \left( \frac{1}{\xP} \right)^{\Delta_\Pom} (1-\beta)^{n(Q^2) + 4}  \;, \\
\label{eq:TriplePomeron}
\chi_3 (s,b,Q^2) &=& \int^{\beta_{max}}_{\beta_{min}} \frac{d \beta}{\beta}\chi_3 (s,b,Q^2,\beta) \;,
\eeq
where, in this case, $\lambda_\Pom(\xP) = R^2_{0 \Pom} + \alpha'_\Pom \ln \left( 1 \big/ \xP \right)$, where $\xP = x\big/ \beta$ is the Bjorken momentum fraction of the pomeron, and the limits of integration are given by $\beta_{min} = x/ {\xP}_{max} = 10 \,x$ and $\beta_{max} = Q^2/(Q^2 + M^2_{min})$. The triple-reggeon eikonal accounts for heavy-mass diffraction, thus $M_{min} = 1$ GeV.\footnote{In the CFSK model \cite{Capella:2000hq}, $\chi_3$ contains a contribution from $\Reg\Pom\Pom$ diagrams (exchanges in the triple-reggeon or pomeron diagrams are
labeled clockwise starting from the left). These are unphysical and have removed in the calculations presented in this paper. Numerically, this contribution is insignificant in calculations of inclusive DIS and are only sizeable in the high-$\xP$, low-$\beta$ region of diffraction which we discuss in detail below.}  A generic diagram of the $\vp p$-interaction is shown in Fig.~\ref{fig:ReggeonDiagram}.

The cross section of the $S$-component has been cast in the standard dipole form
\beq
\sigma_S^{(tot)} (s,Q^2) \,=\, \sum_{T,L} \int^{r_0}_0 dr \, \int^1_0 dz \, \left| \psi^{T,L}(r,z) \right|^2 \sigma_S (r,s,Q^2) \;,
\eeq
where $r_0$ is a cut-off parameter on the size of the dipoles, to be fitted, and $\psi^{T(L)}$ are the wave functions of the $\qqbar$ pair corresponding to transverse and longitudinal polarizations of the virtual photon, the corresponding squares given by
\beq
\left| \psi^T (r,z) \right|^2 \,&=&\, \frac{6 \alpha_{e.m.}}{4 \pi^2} \, \sum_q \, e_q^2 \,\left[ z^2 + (1-z)^2\epsilon^2 K_1^2(\epsilon r) \,+\, m_q^2 K_0^2(\epsilon r) \right] \;, \\
\label{eq:psiL}
\left| \psi^L (r,z) \right|^2 \,&=&\, \frac{6 \alpha_{e.m.}}{4 \pi^2} \, \sum_q \, e^2_q \, \left[ 4 Q^2 z^2 \,+\, (1-z)^2 K_0^2(\epsilon r) \right] \;.
\eeq
Here $\epsilon^2 = z(1-z) Q^2 + m_q^2$, and $K_0$ and $K_1$ are McDonald functions. We use the same quark mass, $m_q$, for all three quark flavors. The dipole-nucleon cross section $\sigma_S(r,s,Q^2)$ can be written in terms of the cross section at a given impact parameter 
\beq
\sigma_S (r,s,Q^2) \,=\, 4 \int d^2b \, \sigma_S (r,b,s,Q^2) \;,
\eeq
where $\sigma_S(r,b,s,Q^2)$ is cast analogously to eqs.~(\ref{eq:LargeComponent}) and (\ref{eq:TotalChi}) for $\sigma_L(b,s,Q^2)$ with $\chi_L$ replaced by $\chi_S$. The dependence of the dipole cross section on $r$ is introduced taking into account that for small dipoles cross sections are proportional to $r^2$, so that the eikonal function is defined as
\beq
\chi_{S 0}^{\Pom} (b,\xi) \,=\, \frac{C_S^{\Pom}\, r^2}{\lambda_{0 \Pom}^S(\xi)} \, \exp \left( \Delta_{\Pom} \xi \,-\, \frac{b^2}{4 \lambda_{0 \Pom}^S(\xi)} \right) \;,
\eeq
where $\lambda_{0 \Pom}^S = R_{0 \Pom S}^2 + \alpha'_\Pom \xi$. Note that secondary ($\Reg$) exchanges do not contribute to the $S$-component.

A notable feature of the model is its growing interaction radius with energy, encoded in the functions $\lambda_{0 k}^L$ and $\lambda_{0 \Pom}^S$, which leads to an increase of the cross section at low $x$. It is also worth noticing that the amount of damping of the functions $\chi_L$ and $\chi_S$, see eq.~(\ref{eq:TotalChi}), are controlled by the parameter $a$ and the function $\chi_3$ which are strongly constrained by diffraction. In this way, unitarization of the cross section and the amount of diffraction are intimately linked.

Several of the model parameters are fixed from studies of the energy dependence of total and diffractive cross sections of hadronic reactions in Regge theory, while the rest were fitted to inclusive and diffractive DIS at low $x$ and $0 < Q^2 < 5$ GeV$^2$ \cite{Capella:2000hq}. A summary of the parameter values is given in Table~\ref{tab:CFSKpar}; the fitted ones from \cite{Capella:2000hq} can be found in the column labeled CFSK. This particular realization of the model conjectured a (non-standard) value of the reggeon intercept, $\Delta_f = -0.3$. We have redone the fit to the same dataset fixing $\Delta_f=-0.5$, which in the following we will denote as CFSK', see rightmost column in Table~\ref{tab:CFSKpar} for details. The new fit works even slightly better than the old one and has several additional advantages which we will describe in detail below. It is interesting to notice that the fitted cut-off on the size of the dipoles in the $S$-component is about $0.2 \div 0.25$ fm.
\begin{table}[t!dp]
\begin{center}
\begin{threeparttable}
\begin{tabular}{c|  c c}
\hline \hline
Fixed parameters & \hspace{0.99cm} CFSK \cite{Capella:2000hq} & CFSK'\\
\hline
\begin{tabular}{c c}
$\Delta_\Pom$ & 0.2 \\
$\Delta_f$ & -0.3 \tnote{a} \\ 
$\alpha'_\Pom$ & 0.25 GeV$^{-2}$ \\
$\alpha'_f$ & 0.9 GeV$^{-2}$ \\
$R_{0 k L}^2$ & 3 GeV$^{-2}$ \\
$R_{0 \Pom S}^2$ & 2 GeV$^{-2}$ \\
$R_{1 k}^2$ & 2.2 GeV$^{-2}$ \\
$\gamma_f$ & 8 \\
$C$ & 1.5
\end{tabular} & 
\begin{tabular}{c c}
$g^2_L(0)$ & $4.56 \times 10^{-3}$ \\
$C_L^f$ & 1.87 GeV$^{-2}$ \\
$C_L^\Pom$ & 0.56 GeV$^{-2}$ \\
$s_0$ & 0.79 GeV$^2$ \\
$a$ & $4.63 \times 10^{-2}$ GeV$^{-2}$ \\
$m_L^2$ & 0.59 GeV$^2$ \\
$C_S$ & 0.185 \\
$r_0$ & 1.06 GeV$^{-1}$ \\
$m^2_q$ & 0.15 GeV$^2$
\end{tabular} &
\begin{tabular}{c}
$5.65 \times 10^{-3}$ \\
2.95 GeV$^{-2} $ \tnote{b} \\
0.46 GeV$^{-2}$ \\
0.79 GeV$^2$ \\
$6.13 \times 10^{-2}$ GeV$^{-2}$ \\
0.70 GeV$^2$ \\
0.105 \\
1.33 GeV$^{-1}$ \\
$1 \times 10^{-3}$ GeV$^2$
\end{tabular} \\
\hline \hline
\end{tabular}
    \begin{tablenotes}
       \item[a] \footnotesize{Put to -0.5 in the CFSK' fit.}
     \end{tablenotes}
\end{threeparttable}
\end{center}
\vspace{-1.em}
\caption{Parameters of the $\vp N$ model: the column labeled CFSK corresponds to the original model parameters, as found in  \cite{Capella:2000hq}; the column labeled CFSK' corresponds to a new fit performed with a smaller value of the reggeon intercept, $\Delta_f$.}
\label{tab:CFSKpar}
\end{table}%

\section{QCD evolution of the inclusive structure function}
\label{sec:QCDinclusive}

In order to generalize the CFSK model to large values of $Q^2$, QCD scaling violations have to be included. The model is therefore used as the initial condition for the DGLAP evolution equations at the initial scale $Q^2_0 = 2$ GeV$^2$ (any  $Q^2_0 \geq 1$ GeV$^2$ can, in principle, be set as the initial scale and the results are rather insensitive to the value of $Q_0$). The initial condition for the evolution equations should be given for all $x$. While the CFSK model is valid only for small values of $x$, it can easily be extended to the $x \sim 1$ region. In order to do so, we follow a standard procedure \cite{Kaidalov:2000wg} multiplying the partonic distributions by the relevant powers of $(1-x)$ as explained below. In what follows we will work at leading order (LO).

Let us first infer the valence quark PDFs from the total cross section. They correspond to the exchange of secondary reggeons. In our model these are only inlcuded in the $L$-component, see eq.~(\ref{eq:LargeComponent}). In order to separate the $\Reg$-contribution, only linear terms in $\chi_L^\Reg$ will be taken since it is sizeable only at not too low $x$, where multi-reggeon exchanges can be neglected. We therefore find that the valence quark contribution to $F_2$ from the model is
\beq
F_{2 \, V}^{low-x} \,=\, \frac{Q^2}{4 \pi^2 \alpha_S} \, 4 g^2_L(Q^2) \, \int d^2b \, \frac{\chi_L^\Reg}{2} \,=\, \frac{2 Q^2}{\pi \alpha_S} \, g^2_L(Q^2) C_L^\Reg \, \xi^{\Delta_\Reg} \;.
\eeq
For the proton, we take $u^{low-x}_V(x,Q^2) / 2 = d^{low-x}_V(x,Q^2)$ \cite{Kaidalov:2000wg} and use
\beq
F_{2 \, V}^{low-x} (x,Q_0^2) = \frac{4}{9} \, x u^{low-x}_V(x,Q^2_0) \,+\, \frac{1}{9} \, x d^{low-x}_V(x,Q^2_0) 
\eeq
as the low-$x$ valence quark distribution at the initial scale. The extension to high-$x$ for the proton is carried out by multiplying the $u_V$-quarks by $(1-x)^{n(Q^2)}$ and the $d_V$-quarks by $(1-x)^{n(Q^2)+1}$, respectively, where
\beq
n(Q^2) \,=\, \frac{3}{2} \, \left(1 + \frac{Q^2}{Q^2 + c} \right) \;,
\eeq
and $c = 3.55$ GeV$^2$ \cite{Kaidalov:2000wg}. Finally, the valence quark distributions are given by
\beq
\label{eq:InitialUv}
x u_V (x,Q_0^2) &=& 2 F_{2 \, V}^{low-x}(x,Q_0^2) \, \left( 1-x\right)^{n(Q_0^2)} \;, \\
\label{eq:InitialDv}
x d_V (x,Q_0^2) &=& F_{2 \, V}^{low-x}(x,Q_0^2) \, \left( 1-x\right)^{n(Q_0^2)+1} \;.
\eeq
The sea quark PDF is given by the sum of the $S$-component and the singlet contribution to the $L$-component, i.e. neglecting all $\Reg$-terms in the latter:
\beq
F_{2 \,Sea}^{low-x} = \frac{Q^2}{4\pi^2 \alpha_S} \, \left(\sigma_S^{(tot)} + \left. \sigma_L^{(tot)}\right|_{C^f_L=0} \right) \;.
\eeq
In order to obtain the PDF's for the different flavors in the sea, we define $S(x,Q^2) \equiv u = \bar{u} = d = \bar{d} = 2 s = 2 \bar{s},$\footnote{Here we neglect the difference between $\bar{u}$ and $\bar{d}$ quarks in the region $x \leq 0.1$.} so that 
\beq
F_{2 \, Sea}^{low-x} (x,Q_0^2) \,=\, \sum_{q,\bar{q}} \, e_q^2 \, x q^{low-x}(x,Q_0^2) \,=\, \frac{11}{9} \, x S^{low-x}(x,Q_0^2) \;.
\eeq
Taking into account the relevant $(1-x)$ factor for the high-$x$ behavior, the sea quark distribution is finally given by
\beq
\label{eq:InitialSea}
x S(x,Q_0^2) = \frac{9}{11} F_{2 \, Sea}^{low-x}(x,Q_0^2) \, (1-x)^{n(Q_0^2)+4} \;,
\eeq
at the initial scale.

At sufficiently large $Q^2$ the term $\chi_S$ can be related to the distribution of gluons in the proton. In the leading logarithmic approximation \cite{Frankfurt:1993it,Frankfurt:1996ri} we have that
\beq
\label{eq:GluonLO}
\sigma_S(r,s,Q^2) \,=\, r^2 \frac{\pi^2}{3} \alpha_S(Q^2) \, x g(x,Q^2) \;,
\eeq
where $x g(x,Q^2)$ is the gluon distribution function of a proton and $\alpha_S$ is the strong coupling constant.  For small values of $\chi_S$, we obtain
\beq
x g^{low-x}(x,Q_0^2) \,=\, \frac{6}{\pi^2 \alpha_S(Q_0^2)} \, \frac{C_S^\Pom}{\lambda_{0 \Pom}^S(\xi)} \int d^2b \; \frac{\exp \left( \Delta_\Pom \xi \,-\, \frac{b^2}{4 \lambda_{0 \Pom}^S(\xi)} \right)}{1 + a \chi_3(b,s,Q^2)} \;.
\eeq
The extrapolation to large values of $x$ is the same as for the sea quarks with an additional $(1-x)^{-1}$ factor \cite{Kaidalov:2000wg}, that is
\beq
\label{eq:InitialGluon}
x g(x,Q_0^2) \,=\, x g^{low-x}(x,Q_0^2) \, (1-x)^{n(Q_0^2) + 3} \;.
\eeq
Thus, the equations (\ref{eq:InitialUv})-(\ref{eq:InitialDv}), (\ref{eq:InitialSea}) and (\ref{eq:InitialGluon}) constitute the initial conditions for the DGLAP evolution equations for the inclusive structure function.

The treatment of heavy quarks is done in the zero-mass variable flavor number scheme. This scheme provides matching prescriptions between $N_f$ and $N_f+1$ evolved PDFs at a given threshold scale, $\mu_T$, proportional to the heavy quark mass, $m_Q$. The proportionality constant is not known theoretically but can be estimated requiring smoothness of observables, see \cite{Haakman:1997kh,Thorne:2008xf}. We take the charm quark mass to be $m_c = 1.4$ GeV, neglect bottom and top contributions and take $\mu_T = 2.5 \, m_c$. Further investigation of the impact of heavy quarks and a detailed comparison to heavy quark structure functions, $F_2^c$ and $F_2^b$, lie beyond the scope of the present work.

\begin{figure}[t!]
\begin{center}
\includegraphics[width=0.5\textwidth]{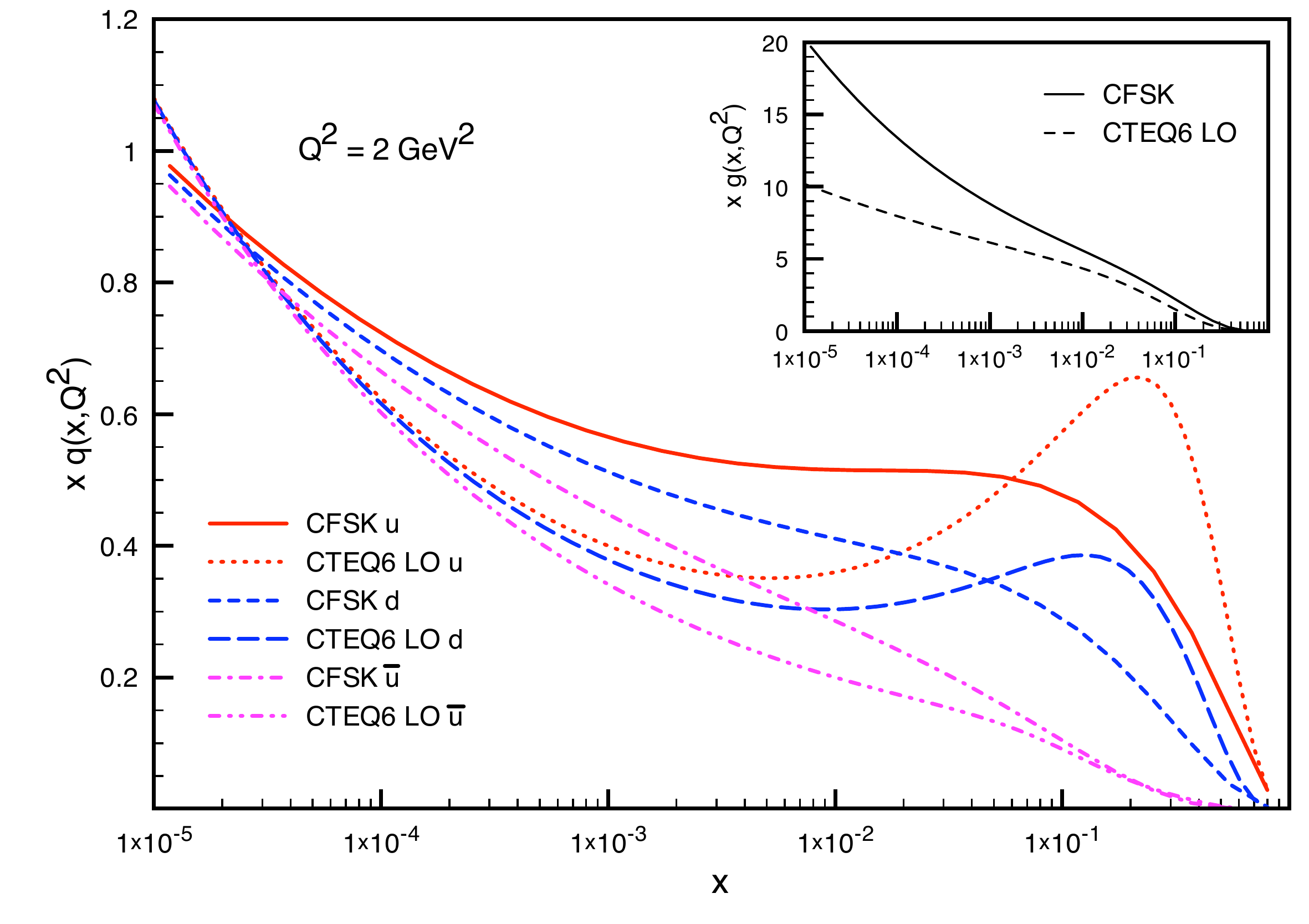}%
\includegraphics[width=0.5\textwidth]{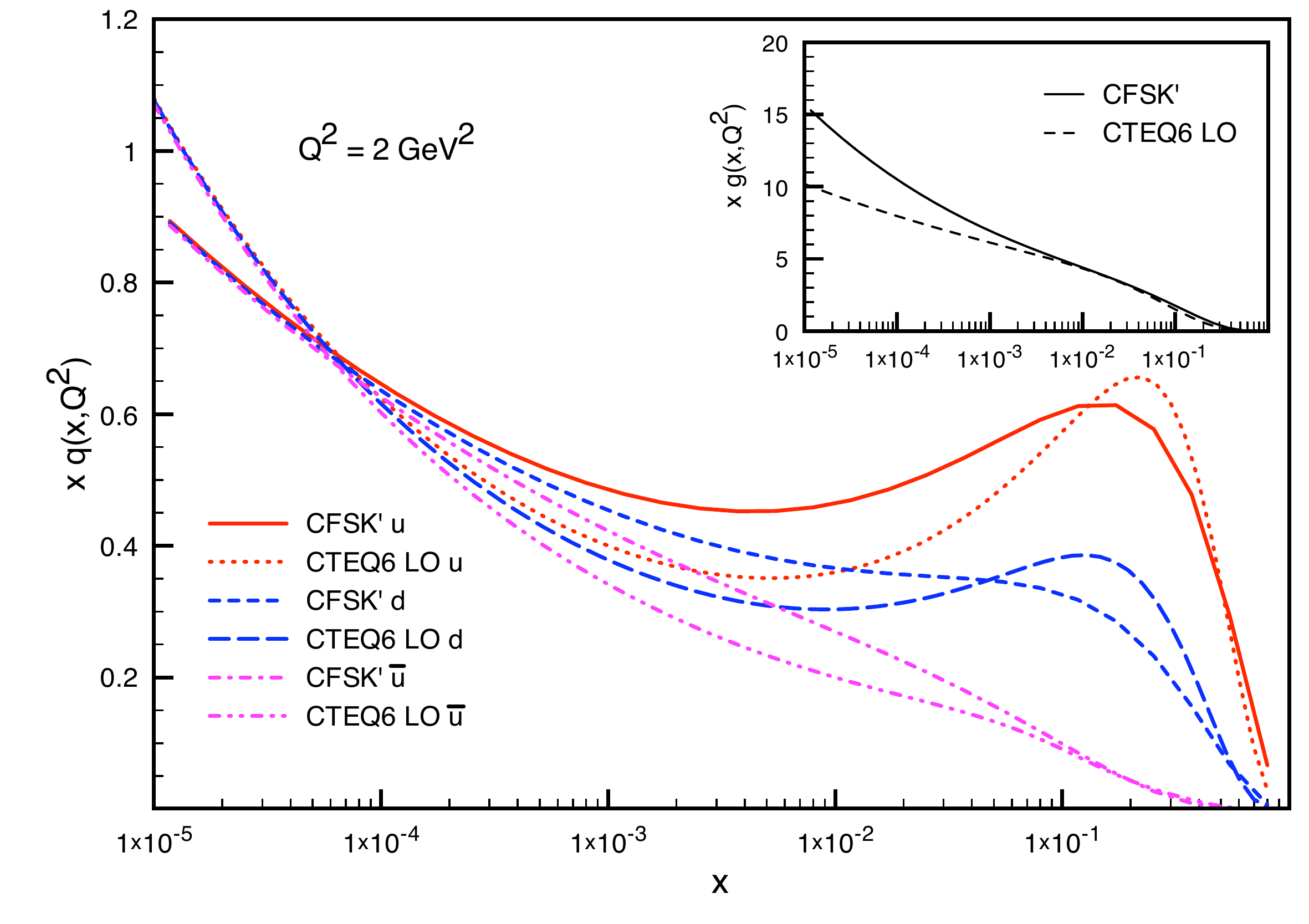}
\caption{Initial  quark and gluon parton distribution functions in CFSK (left) and CFSK' (right) models compared to the CTEQ6 LO parameterization \cite{Nadolsky:2008zw} at $Q_0^2 = 2$ GeV$^2$.}
\label{fig:InitPDF}
\end{center}
\end{figure}
The parton distribution functions at the initial scale at leading order are thus fixed unambiguously as described above and they have to fulfill the valence and momentum sum rules, namely
\beq
\begin{array}{l} 
\begin{array}{c}	
\label{eq:ValenceSum}
\int dx \; u_V(x,Q_0^2) \,=\, 2 \;, \\
\int dx \; d_V(x,Q^2_0) \,=\, 1 \;,
\end{array} \\
\label{eq:MomentumSum}
\int dx \, x \Big( u_V(x,Q^2_0) \,+\, d_V(x,Q^2_0) \,+\, 5 S(x,Q^2_0) \,+\, g(x,Q^2_0) \Big) \,=\, 1 \;.
\end{array}
\eeq
The partonic decomposition of the original CFSK model does not automatically fulfill these rules, which would force us to introduce uncomfortably large overall normalization factors. The CFSK' fit described above, on the other hand, automatically fulfills the sum rules to a good approximation. We have verified that eqs.~(\ref{eq:ValenceSum}) are satisified with an accuracy better that $5 \div 10$\%. To further improve the sum rules we have increased the parameter $C^f_L$ by 12\%, see Table~\ref{tab:CFSKpar}. The fraction of the proton momentum carried by the gluons at the initial scale is then 
\beq
\frac{\int dx \, xg(x,Q_0^2)}{\int dx \, x  \left( S(x,Q_0^2) + g(x,Q_0^2) \right)} \,=\, \left\{ \begin{tabular}{cl} 0.59 & \mbox{   in CFSK  ,} \\ 0.49 & \mbox{   in CFSK'  .} \end{tabular} \right.
\eeq

The initial parton distribution functions (PDFs) for both fits are compared to the CTEQ6 LO parameterization \cite{Nadolsky:2008zw} in Fig.~\ref{fig:InitPDF}. The choice of $\Delta_f=-0.5$ is clearly improving the CFSK' valence PDF contribution, and the smaller value of the $C_S$ parameter than in the original CFSK brings the gluons closer to the fit of CTEQ6. Since the QCD sum rules seem to be better satisfied in the CFSK' model with no additional normalization of the input PDFs, we will continue using this set of parameters in what follows.

\begin{figure}[t!]
\begin{center}
\includegraphics[width=0.9\textwidth]{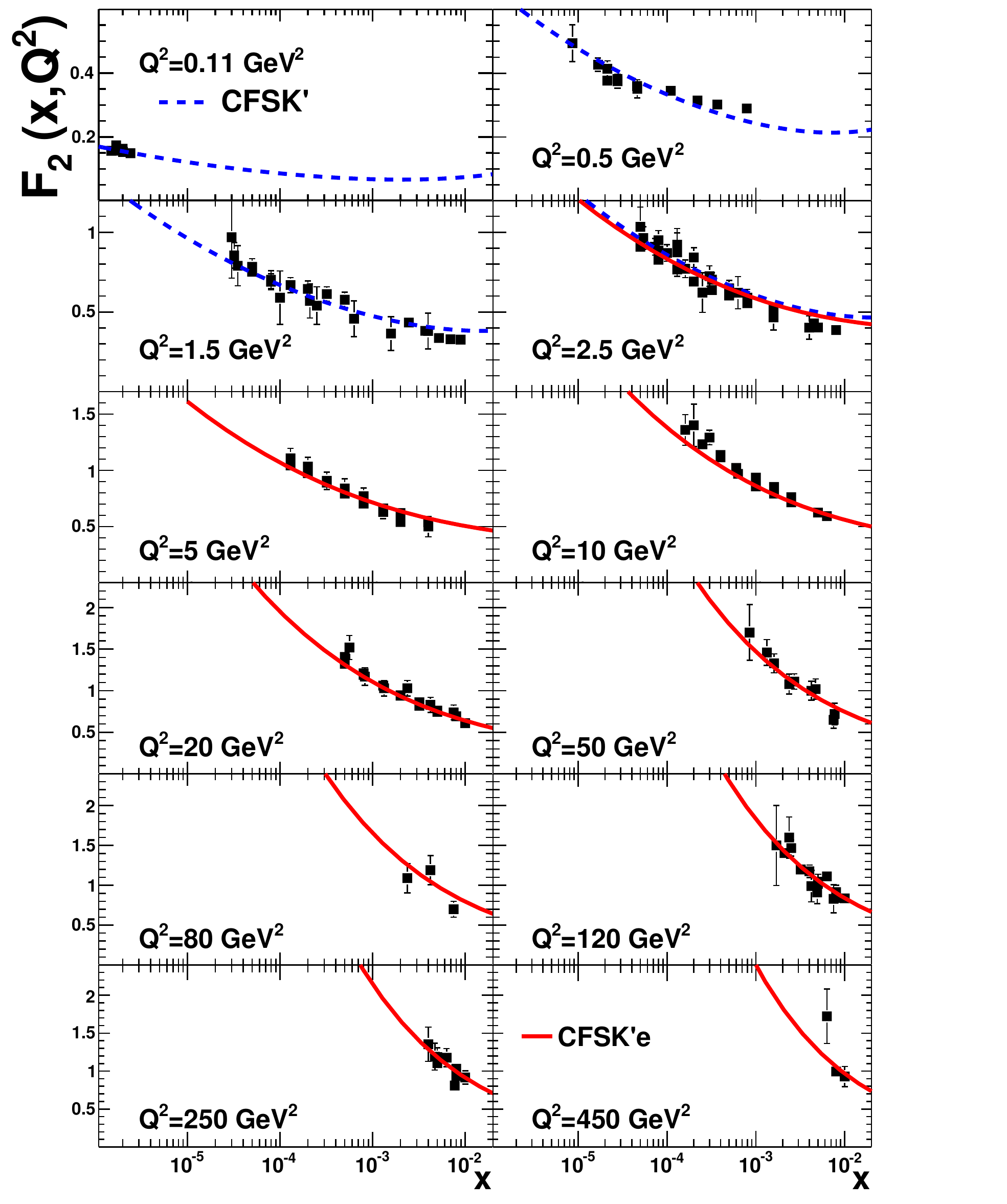} 
\caption{The structure function $F_2$ in the CFSK'e model compared to data. The blue curve is the unevolved CFSK' model and the red solid curves is the CFSK' evolved at LO. Black points are experimental data \cite{Adams:1996gu,Arneodo:1996qe,Abt:1993cb,Ahmed:1995fd,Aid:1996au,Adloff:1997mf,Adloff:1999ah,Adloff:2000qk,Derrick:1993fta,Derrick:1994sz,Derrick:1995ef,Derrick:1996hn,Breitweg:1997hz,Breitweg:1998dz,Breitweg:2000yn,Chekanov:2001qu}.}
\label{fig:F2evolution}
\end{center}
\end{figure}
We have subsequently performed QCD evolution at LO, with $\alpha_{S} (M_Z^2) = 0.126$ using the QCDNUM evolution code \cite{Botje:1999dj}. The $F_2$, and the associated parton distribution functions, thus obtained are denoted CFSK'e (evolved). Comparing with data from experiments \cite{Adams:1996gu,Arneodo:1996qe,Abt:1993cb,Ahmed:1995fd,Aid:1996au,Adloff:1997mf,Adloff:1999ah,Adloff:2000qk,Derrick:1993fta,Derrick:1994sz,Derrick:1995ef,Derrick:1996hn,Breitweg:1997hz,Breitweg:1998dz,Breitweg:2000yn,Chekanov:2001qu}, a total of 847 data-points, the resulting $\chi^2 \big/$d.o.f. is 2 both for CFSK' ($0 < Q^2 \leq 2$ GeV$^2$) and in the evolved case ($2 < Q^2 < 1000$ GeV$^2$).\footnote{Here, we have not considered the normalization errors of the data sets when calculating the $\chi^2$.} The results of CFSK' and CFSK'e are shown and compared to a set of the experimental data in Fig.~\ref{fig:F2evolution}. Summarizing, we have shown that the proposed model is in good agreement with experimental data in the whole range of accessible $Q^2$ and down to very low $x$.

\begin{figure}[t!]
\begin{center}
\includegraphics[width=0.9\textwidth]{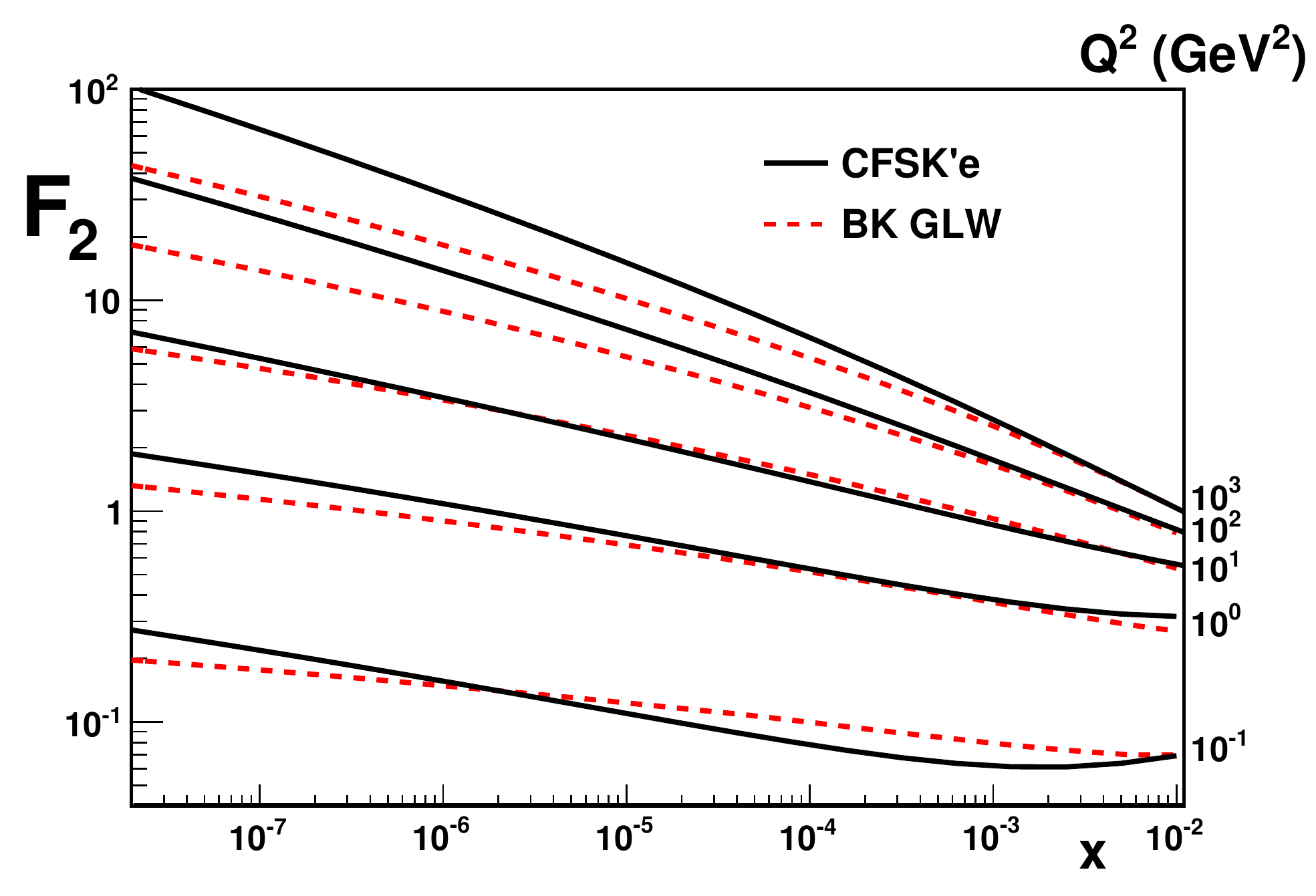} 
\caption{Predictions for $F_2$ in the CFSK'e model and from a numerical solution of the BK equation\cite{Albacete:2009fh}.}
\label{fig:F2evolutionBK}
\end{center}
\end{figure}
The LO DGLAP evolution takes into account large logarithms in $Q^2$, yet at very high energies one expects large corrections arising from the missing terms in $\ln (1 \big/ x)$ which are more properly accounted for in the BK equation or in linear resummation schemes \cite{Altarelli:2008aj,Ciafaloni:2007gf}. In Fig.~\ref{fig:F2evolutionBK} we compare predictions of low-$x$ $F_2$ to the recent solution of the BK equation presented in \cite{Albacete:2009fh}, where an initial $\qqbar N$ cross section in the form of the GBW model was assumed. We notice large differences between the models at very large $Q^2$, which may partly be caused by the fact that the calculation presented in \cite{Albacete:2009fh} does not take into account heavy flavors which are important at those momentum scales and partly by the absence of impact parameter dependence in the calculation \cite{Albacete:2009fh}. Regardless of this, the deviations between the models for $x \leq 10^{-6}$ at $Q^2 > 10$ GeV$^2$ hint of a breakdown of the DGLAP equations. Although the CFSK model satisfies unitarity, the observed difference is purely an effect of high-energy QCD and will be extremely interesting to study in the future. 

\section{The longitudinal structure function}
\label{sec:FL}

The measurement of the longitudinal structure function of the proton at HERA has been eagerly awaited. The longitudinal structure function is zero at leading order QCD and is a direct measure of the size of the gluon distribution which had before this only been accessed through the scaling violations of the total structure function. Thus, it is believed that it can be a sensitive probe to saturation effects at low $x$. In the dipole model at leading logarithmic accuracy $F_L$ only probes the dynamics of small, perturbative configurations.

Although the longitudinal structure function is a NLO observable in the QCD improved parton model within collinear factorization, one can calculate this quantity in the framework of the dipole model using the connection between the dipole-nucleon cross section and the gluon distribution function at LLA.  On the one hand, we can use directly the form of the dipole-nucleon cross section of the unevolved CFSK' model, eq.~(\ref{eq:InitialGluon}) and on the other hand we can make use of the relation of this cross section to the gluon distribution at $Q^2 \geq Q^2_0$, eq.~(\ref{eq:GluonLO}), which we extract from the perturbative calculation, CFSK'e.

Note that in the latter case there is a slight mismatch between the non-perturbative and perturbative gluons due to the integration over impact parameter. In principle, the definition in eq.~(\ref{eq:InitialGluon}) gives us the first term in a series of the (impact parameter dependent) quasi-eikonal model. To be able to resum the quasi-eikonal series we should include the correct impact parameter dependence of each individual term. We therefore define the impact parameter dependent gluon density in full analogy with each of the pomeron terms as follows
\beq
x g(b,x,Q^2) = \frac{\exp \left\{ -b^2 \Big/ 4 \lambda_{0 \Pom}^S (\xi)\right\}}{4 \pi  \lambda_{0 \Pom}^S (\xi)} \, x g(x,Q^2) \;,
\eeq
where we have ensured the proper normalization. The final $\qqbar-N$ cross section is therefore
\beq
\label{eq:FLsigma}
\sigma_S(r,x,Q^2) = 4 \int d^2b \, \frac{1}{2 C} \left[ 1 - \exp \left\{ -C \frac{\pi^2 \alpha_S(Q^2)}{6} \frac{\exp \left\{ -b^2 \Big/ 4 \lambda_{0 \Pom}^S (\xi)\right\}}{4 \pi  \lambda_{0 \Pom}^S (\xi)} \, x g(x,Q^2) \, r^2 \right\} \right] \;,
\eeq
where both the gluon distribution and the strong coupling constant are calculated at LO. The resulting longitudinal structure function is found by convoluting $\sigma_S$ in eq.~(\ref{eq:FLsigma}) with the wave-function for longitudinally polarized photons in eq.~(\ref{eq:psiL}).
\begin{figure}[t!]
\begin{center}
\includegraphics[width=0.48\textwidth]{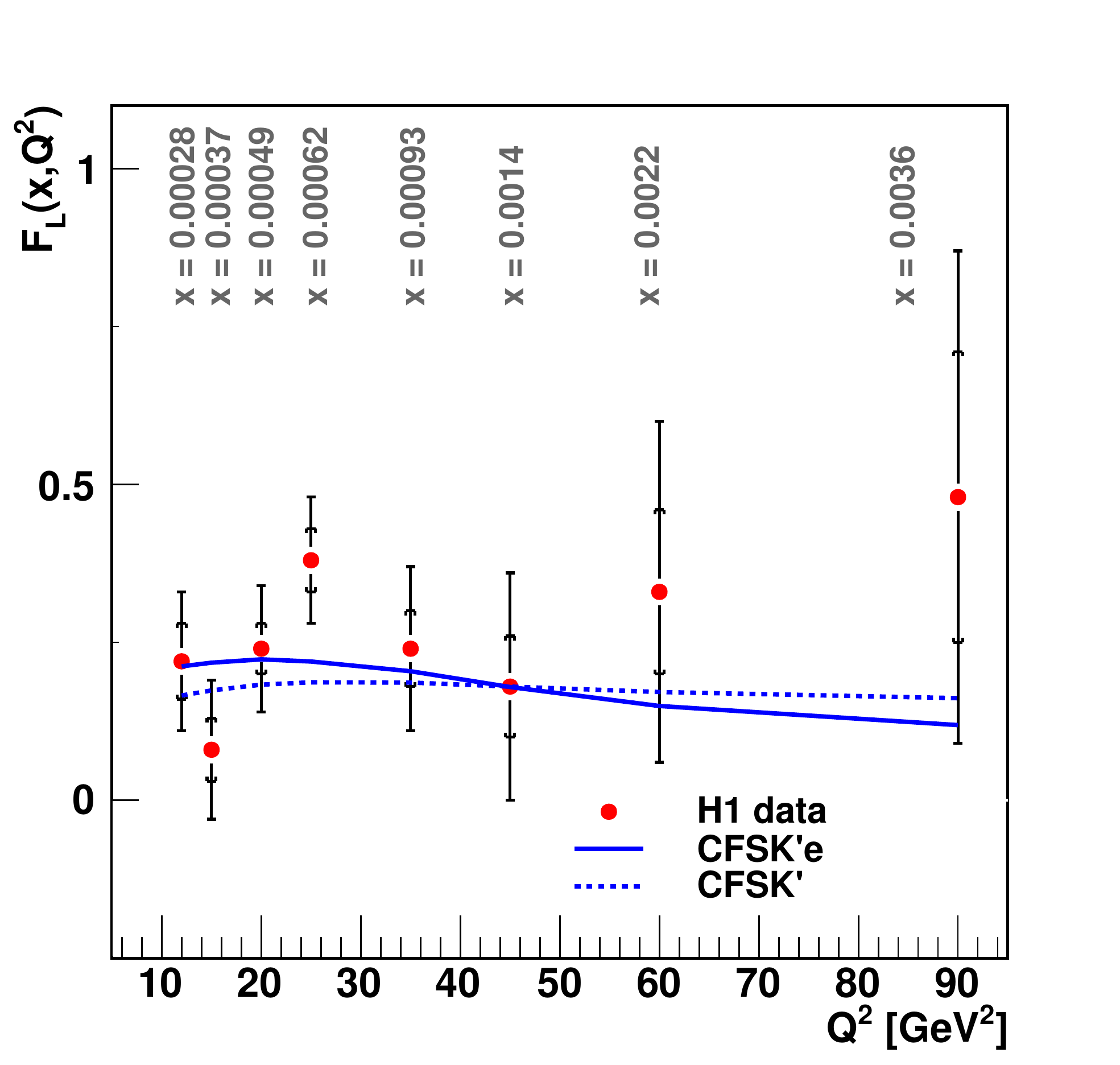}
\includegraphics[width=0.48\textwidth]{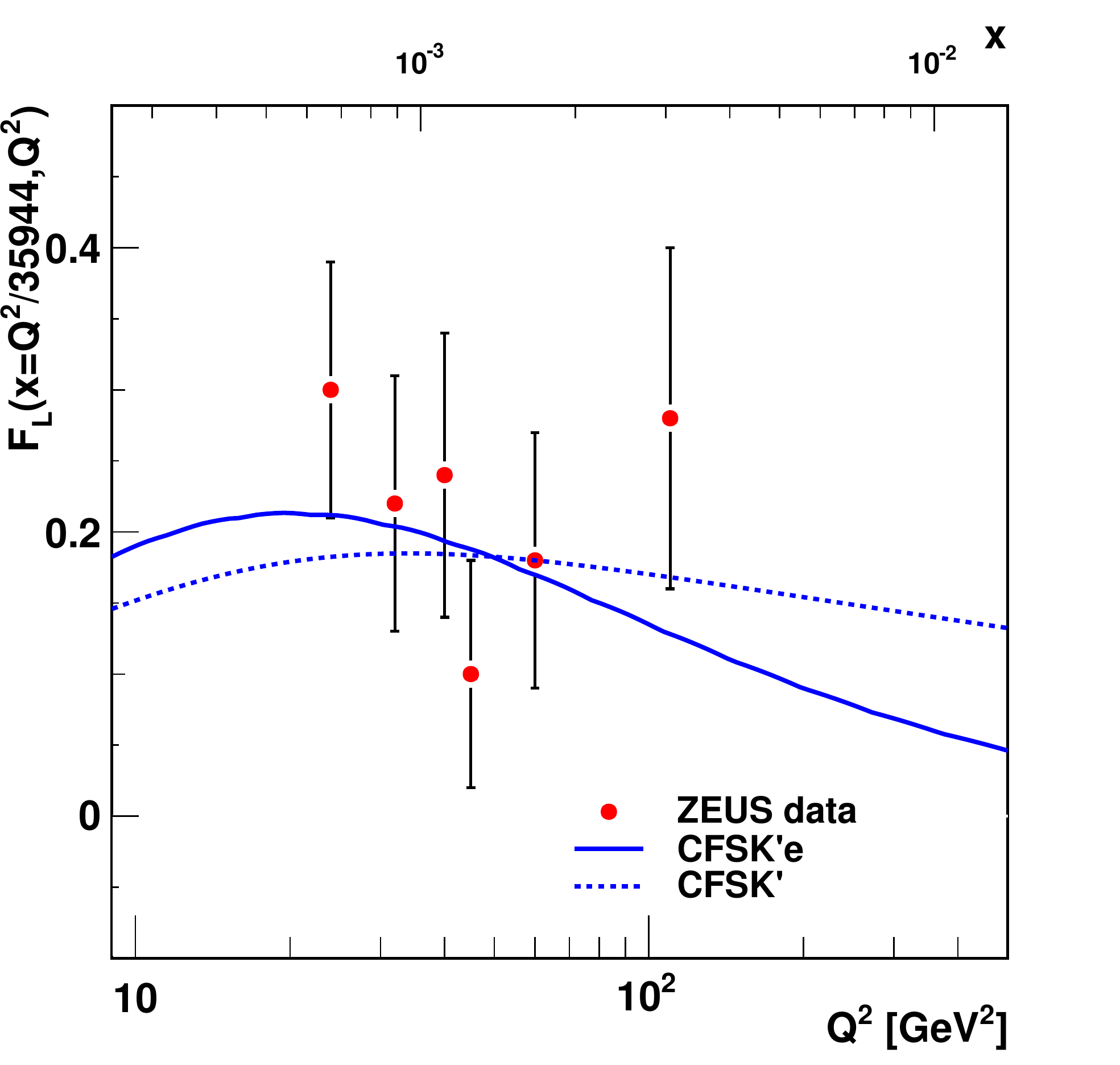}
\caption{The longitudinal structure function of the proton as a function of $Q^2$ in CFSK' and CFSK'e compared to data averaged over $x$ from H1 \cite{Aaron:2008tx} (left figure) and ZEUS \cite{Chekanov:2009na} (right figure).}
\label{fig:CFSK_FL1}
\end{center}
\end{figure}

\begin{figure}[t!]
\begin{center}
\includegraphics[width=0.7\textwidth]{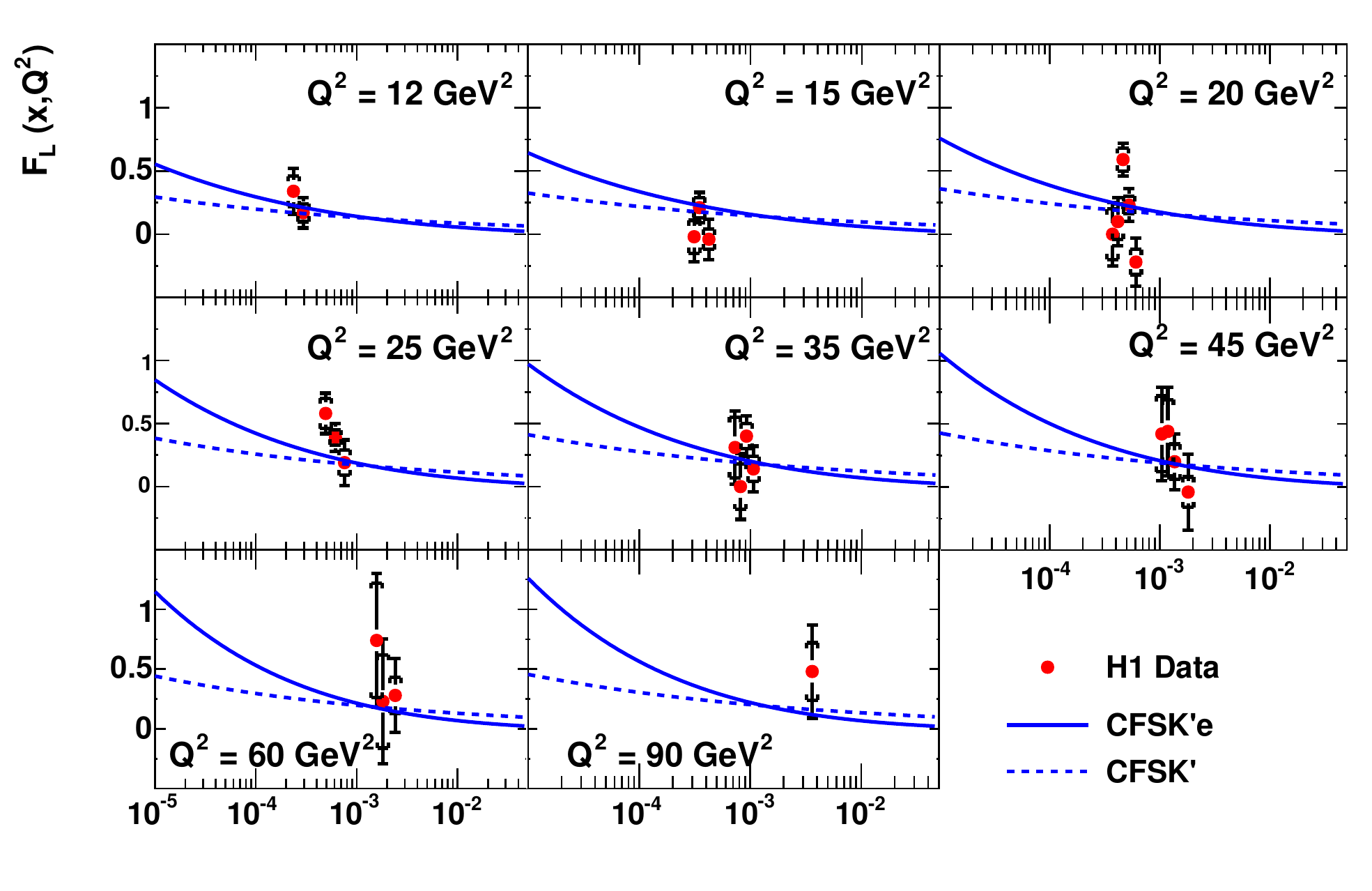}
\caption{The $x$-dependence of the longitudinal structure function of the proton for several $Q^2$ bins in the CFSK' and CFSK'e models compared to H1 data \cite{Aaron:2008tx}.}
\label{fig:CFSK_FL2}
\end{center}
\end{figure}

We compare the two prescriptions for the dipole-nucleon cross section described above to H1 \cite{Aaron:2008tx} and ZEUS \cite{Chekanov:2009na} data in Figs.~\ref{fig:CFSK_FL1} and \ref{fig:CFSK_FL2}. At the present moment the experimental data have too large errors to distinguish between models, but more precise data at low-$Q^2$ could increase the discriminating power of the observable. 

\begin{figure}[t!]
\begin{center}
\includegraphics[width=0.9\textwidth]{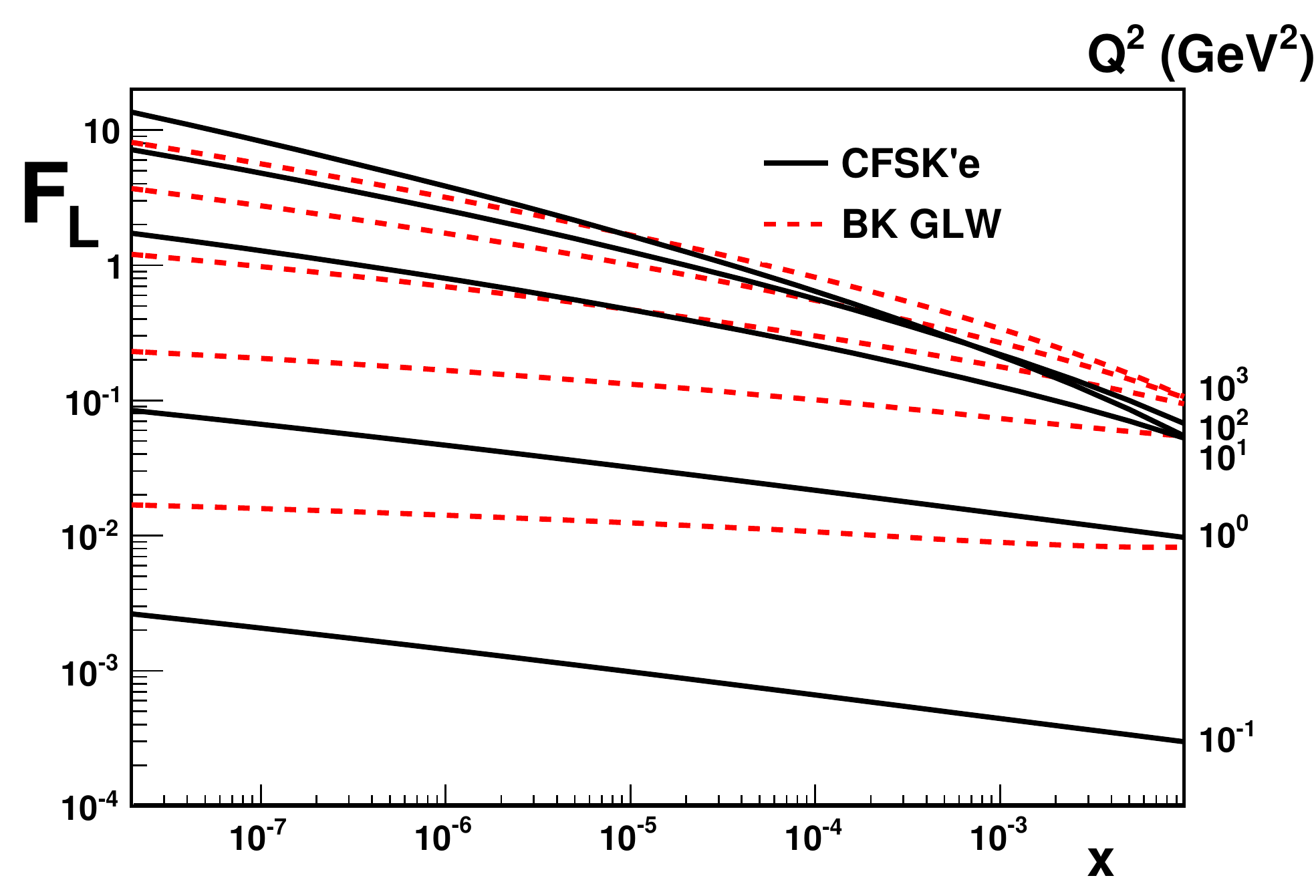} 
\caption{Predictions for $F_L$ in the CFSK model and from a numerical solution of the BK equation\cite{Albacete:2009fh}. Note that the lower red, dashed curve of the BK solution corresponds to $Q^2 = 10^{-1}$ GeV$^2$.}
\label{fig:FLevolutionBK}
\end{center}
\end{figure}

Finally, $F_L$ at low $x$ is expected to be sensitive to saturation effects which could be probed at a future electron-proton collider. The authors of \cite{Rojo:2009ut} found that combined data on $F_2$ and $F_L$ give a strong discriminating power in revealing saturation effects. In particular, we note an order of magnitude difference between the prediction of the CFSK model and the BK model at low $x$ and $Q^2$, which are compared in Fig.~\ref{fig:FLevolutionBK}. This  region lies beyond the reach of perturbation theory and can shed light on the transition to the non-perturbative regime. 

\section{QCD evolution of the diffractive structure function \label{sec:QCDdiffractive}}

Using the AGK cutting rules \cite{Abramovsky:1973fm} we can readily obtain the diffractive cross section from the formulas of the total cross section described above. Once more, the diffractive cross section will consist of contributions from the large and small partonic configurations of the virtual photon wave-function, while the leading contribution arise only from the former, unlike in the inclusive case. Therefore, the CFSK model includes also the explicit contribution from $3\Pom$ diagrams which are responsible for high-mass diffraction (the low-$\beta$ region). For specific details on the diffractive part of the CFSK model we refer the reader to \cite{Capella:2000pe,Capella:2000hq} and Appendix~\ref{app:CFSKdiffractive}.

The collinear factorization theorem of the diffractive cross section, $\xP \sigma^{\D (4)} (\beta,\xP,Q^2,t)$, is valid at fixed $\xP$ and $t$ only for the resolved photon \cite{Collins:1997sr}. Nonetheless, experimental data show to a good approximation that diffractive DIS data satisfy proton vertex factorization, whereby the dependences on variables which describe the scattered proton ($\xP,t$) factorize from those describing the hard partonic interaction ($Q^2,\beta$). This property is also known as Regge factorization  \cite{Ingelman:1984ns}. For example, the slope parameter $B$, extracted by fitting the $t$ distribution to the form $d \sigma / d t \propto e^{B \,t} $, shows no significant variations from the average value \cite{Chekanov:2008fh}. Also, one observes no significant variation of the pomeron intercept with $Q^2$ \cite{Chekanov:2008fh}.

These observations hold as long as one-pomeron exchange dominates the cross section. The CFSK model involves more complicated diagrams and respects this naive factorization only in a limited kinematical region. On one hand, the multi-reggeon exchanges change the effective intercept of the pomeron. On the other hand, the low-$\beta$ and high-$\xP$ region is dominated by reggeon exchange, a missing piece in the original formulation of CFSK. For the sake of simplicity, comparing our calculations to the experimentally available $t$-integrated (reduced) diffractive cross section, $\xP \sigma_r^{\D(3)} (\beta,\xP,Q^2)$, we will make use of the general Regge factorization, representing the diffractive structure functions as
\beq
\label{eq:ReggeFactorization}
F_{2 \D}^{(3)}(\xP, \beta,Q^2) \,=\, \overline{f}_{\Pom/p} (\xP) \, F_2^\Pom (\beta,Q^2) + n_{\Reg} \overline{f}_{\Reg/p} (\xP) \, F_2^{\Reg} (\beta,Q^2) \;.
\eeq
where $F_2^i$ are the reggeon/pomeron structure functions and $\overline{f}_{k/p}$ is the $t$-integrated ($k=\Reg_i,\Pom$) flux factor of the proton
\beq
\label{eq:PomFlux}
\overline{f}_{k/p} \,=\, \int_{t_{cut}}^{t_{min}} dt  \; \frac{A_k}{\xP^{2 \alpha_k(t) -1}} e^{B_i t} \;,
\eeq
where $t_{min}=m_N^2 \xP^2$ and $t_{cut} = -$1 GeV$^2$. Finally, $n_\Reg$ is an unknown normalization of the reggeon contribution.

\begin{figure}[t!]
\begin{center}
\includegraphics[width=0.5\textwidth]{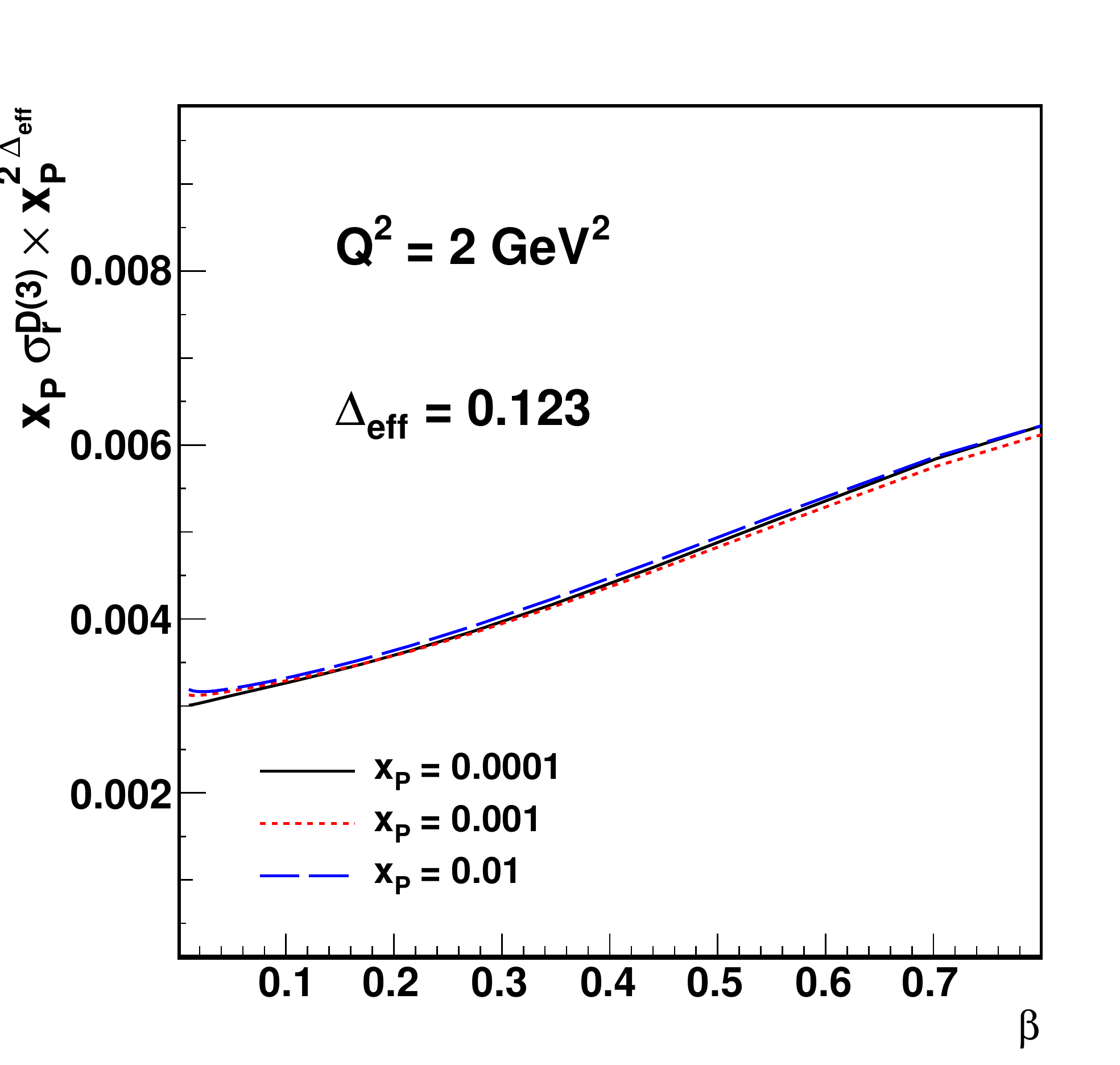}
\caption{Regge factorization in the CFSK model.}
\label{fig:ReggeFactorization}
\end{center}
\end{figure}

Concentrating on the pomeron contribution to $F_{2 \D}^{(3)}$ for the time being, we have checked numerically that the CFSK model of diffraction follows a universal trend, such that $\xP F_{2 \D}^{(3)} \times \xP^{2 \Delta_{eff}} \approx f(\beta)$ for a large range of $\xP$ at the inital scale $Q_0^2$, where $\Delta_{eff}=1-\alpha_\Pom^{eff}(0)=0.123$ is the effective pomeron slope, see Fig.~\ref{fig:ReggeFactorization} for details. Thus, Regge factorization is very well satisfied in the CFSK model, a slight breakdown observed at low $\beta$ values and $\xP \geq 10^{-2}$. 

The pure multi-pomeron contribution at a given ${\xP}_0$ can consequently be defined as
\beq
\left[ \xP \FD \right]_0 \,=\, \left. \xP F_{2 \D}^{(3)}(\xP,\beta,Q^2)\right|_{{\xP}_0} \, = \, \tilde{A}_\Pom \, 
{\xP}_0^{-2 \Delta_{eff}} \, F_2^\Pom (\beta,Q^2) \;,
\eeq
where $ \tilde{A}_\Pom$ is an overall normalization and the $\Reg \Pom \Pom$ contribution is not included. We will hereafter choose ${\xP}_0 = 0.01$ as a reference. Thus, the value of the diffractive structure function at a given $\xP$ is simply
\beq
\label{eq:newF2D3}
\xP \FD(\xP,\beta,Q^2) \,=\, \left( \frac{{\xP}_0}{\xP} \right)^{2 \Delta_{eff}} \, \left[ \xP \FD \right]_0 \;.
\eeq
We carry on with an attempt of a partonic decomposition of the pomeron structure function, analogously to the inclusive case. 

The normalization of the pomeron flux factor, $\tilde{A}_\Pom$, is unknown and therefore we define the singlet quark diffractive parton distribution at the initial scale as
\beq
\label{eq:singdPDF}
\left[ \beta \tilde{S}^\Pom(\beta,Q_0^2) \right]_0 \, = \, \frac{9}{11} \left[ \xP \FD(\xP,\beta, Q^2_0) \right]_0 \;,
\eeq
where we have assumed three active quarks flavors and suppressed the strange quarks by a factor 2 in analogy to the proton PDFs explained in Sec.~\ref{sec:QCDinclusive}. The brackets $\left[ ... \right]_0$ denote that this value is taken at a given ${\xP}_0$. The general $\xP$-dependence is given through the relation in eq.~(\ref{eq:newF2D3}).

Concerning the gluons, since we are not able to separate the flux and the diffractive PDFs (dPDFs), we effectively obtain the $\beta$ and $Q^2$ dependences through the QCD evolution and gain access to a product of the Pomeron flux 
and gluon dPDF given by
\beq
\label{eq:PomGluon}
\xP \overline{f}_{\Pom/p}(\xP) \, \beta g^\Pom(\beta,Q^2) \,=\, \left( \frac{{\xP}_0}{\xP} \right)^{2 \Delta_{eff}} 
\left[ \beta \tilde{g}^\Pom(\beta,Q^2) \right]_0 \;,
\eeq
where the normalization is the same for gluons and quarks. There is no {\it a priori} procedure to extract the gluon distribution from the CFSK model in the diffractive case. Thus, the last factor on the right hand side of eq.~(\ref{eq:PomGluon}) is parameterized as
\beq
\left[ \beta \tilde{g}^\Pom(\beta,Q^2) \right]_0 = A_g \beta^{B_g} (1-\beta)^{C_g} (1 + D_g \sqrt{x}) \exp \left\{ - \frac{0.001}{1-x} \right\} \;,
\label{eq:GluonFit}
\eeq
so that it is integrable. Since, in the single-pomeron exchange model, the ratio of gluon and sea-quark distributions in the proton should equal the corresponding ratio in the pomeron at low $x$ and $\beta$, respectively, we fix the parameter $B_g$ to the standard value, $B_g = - \Delta_\Pom$. The remaining parameters, $A_g$, $C_g$ and $D_g$, of the gluon dPDF at the initial scale have to be extracted from the data.

The description of diffractive data ($Q^2 > 2$ GeV$^2$) in the whole region of $\xP$ and $\beta$ demands a careful analysis of the necessary components \cite{Kaidalov:1979jz}. As shown in \cite{Adloff:1997mi,GolecBiernat:1997vy}, the inclusion of reggeon terms is crucial for describing data at high $\xP$. A missing piece in the original model is the $\Reg \Pom \Reg$ contribution which is dominant in the triple-reggeon region, i.e. at low $\beta$ and high $\xP$. Its diffractive cut corresponds to a reggeon-exchange with large-mass diffraction. One should accordingly introduce parton densities in the reggeon and, following the standard procedure \cite{Aktas:2006hy}, we identify them with the pion ones; together with the standard expression for the reggeon flux, see eq.~(\ref{eq:PomFlux}), we thus obtain the reggeon contribution to $\FD$. In the present work we have employed the LO DGLAP fit from \cite{Gluck:1999xe}, which was shown to be in reasonable agreement with recent leading neutron data from HERA \cite{Collaboration:2010ze}. The parameters of the reggeon trajectory and $t$-slope are the experimentally extracted ones, listed in \cite{Aktas:2006hy}. Finally, the normalization of the flux was set so that $\xP \int^{t_{min}}_{t_{cut}} f_{\Reg/p} dt = 1$ at $\xP = {\xP}_0$. 

We have performed a LO DGLAP fit (with $\alpha_S(M_Z^2)$ and $\mu_T$ as described in Sec.~\ref{sec:QCDinclusive}) of the gluon dPDF parameters and $n_\Reg$ of the reggeon contribution together with the input sea dPDF given in eq.~(\ref{eq:singdPDF}) to experimental data on the diffractive structure function from H1 \cite{Aktas:2006hy} (LRG, 461 points) and ZEUS \cite{Chekanov:2008fh} (LRG, 277 points and LPS, 118 points). 
\begin{table}[t]
\begin{center}
\begin{tabular}{l| c |c |c |c}
 & $A_g$ & $C_g$ & $D_g$ & $n_\Reg$ \\
 \hline
 fit result & 0.108 & -1.82 & -0.91 & 0.0107 \\
\end{tabular}
\end{center}
\caption{Parameters of the gluon density of the pomeron and the normalization of the reggeon contribution obtained from a fit to diffractive data \cite{Aktas:2006hy,Chekanov:2008fh}.}
\label{tab:GluonDiffPar}
\end{table}%
The resulting $\chi^2\big/\mbox{d.o.f.}$ for both the unevolved and evolved CFSK' models with $n_\Reg=0$ are large. The largest deviations arise at high $\xP$ where precise data dictate a more careful treatment of the reggeon contribution to diffraction. Leaving $n_\Reg$ as a free parameter we obtain a $\chi^2\big/\mbox{d.o.f.}$ of 1.8, improving the agreement significantly. The resulting values of the parameters obtained from the fit are listed in Table~\ref{tab:GluonDiffPar}.

\begin{figure}[tp]
\begin{center}
\includegraphics[width=0.95\textwidth]{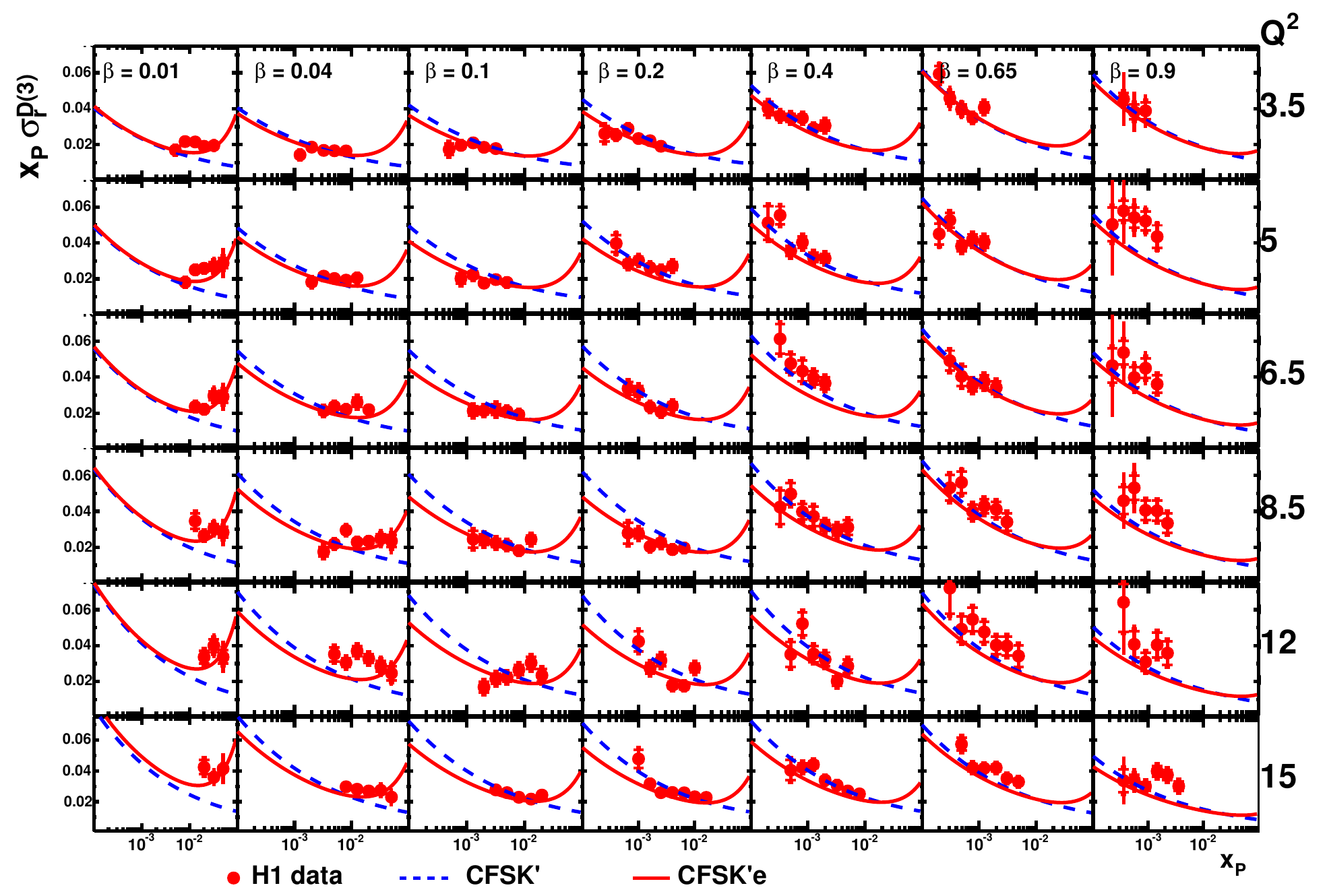}
\caption{QCD evolution at LO of the CFSK' initial condition at $Q^2_0 = 2$ GeV$^2$ as described in the text compared to a set of the H1 data  \cite{Aktas:2006hy}.}
\label{fig:DiffEvolMF}
\end{center}
\end{figure}

The QCD evolved CFSK' initial condition with the fitted parameterization of the gluon dPDF and the reggeon contribution are compared to the H1 LRG data for the lowest $Q^2$ bins in Fig.~\ref{fig:DiffEvolMF} together with the unevolved CFSK' model. The overall  description of the data seems satisfactory in both models, except the low-$\beta$ bins where the missing reggeon contributions leads to larger deviations for the model without evolution.
The reason for the fairly good description of data at high $Q^2$ in the original model can, in fact, be traced back to the dipole form of the small component of the 3$\Pom$ contribution which gives rise to a logarithmic growth in $Q^2$. 

Whereas the description of the inclusive proton $F_2$ at $Q^2 > 2$ GeV$^2$ calls for a extension of the CFSK model while the scaling violations of the diffractive cross section are much better accounted for, we have presented, for the sake of consistency, a complete procedure for a partonic decomposition of the original model and subsequent QCD evolution for both the inclusive and diffractive calculations, dubbed CFSK'e. The proper magnitude of the scaling violations of the diffractive part is crucial for the comparison with the inclusive $F_2$ in the calculation of nuclear shadowing in Glauber-Gribov theory \cite{Gribov:1968gs}, giving rise to the correct $Q^2$-evolution of shadowing.

\begin{figure}[tp]
\begin{center}
\includegraphics[width=0.4\textwidth]{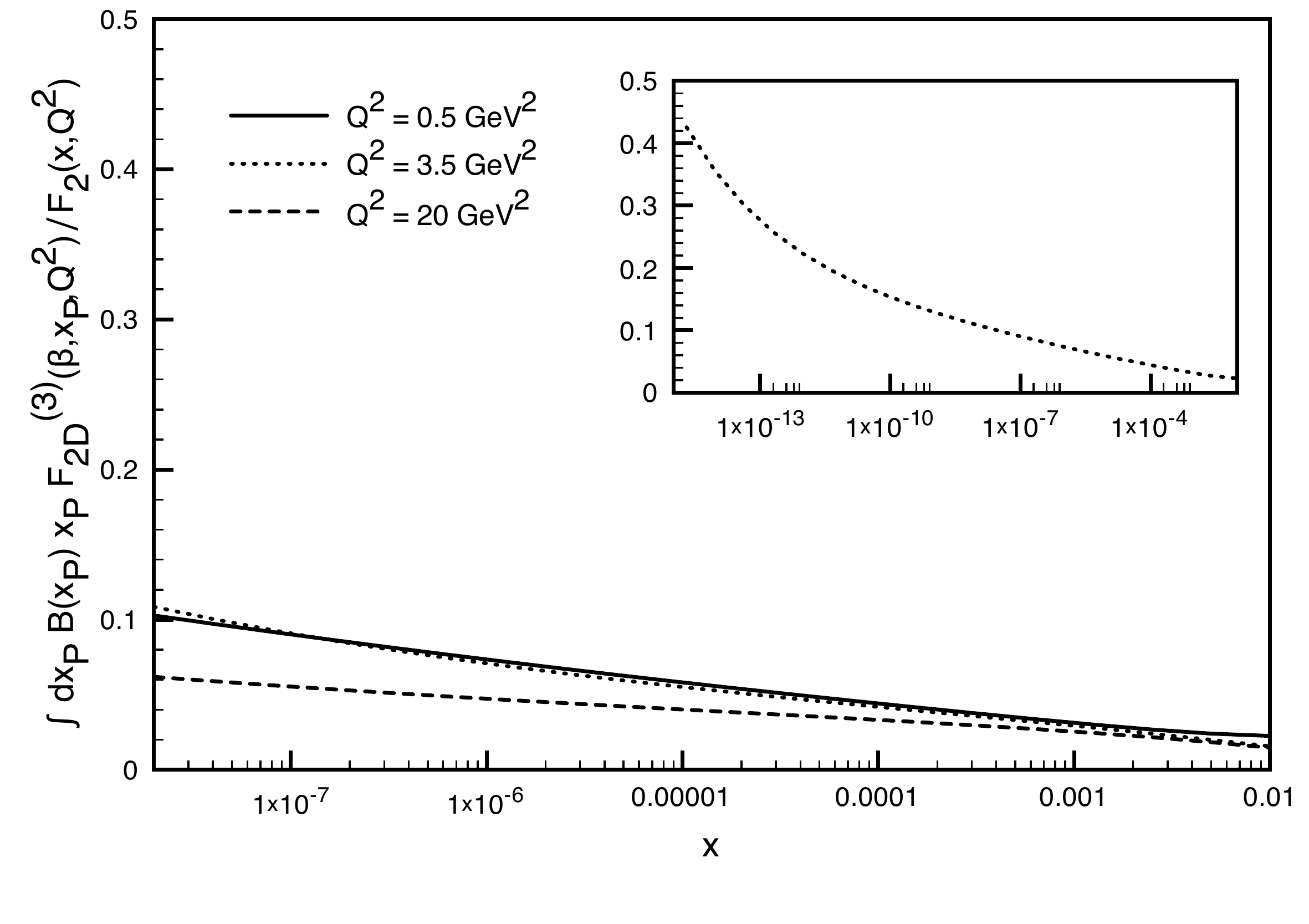}
\caption{The Pumplin ratio for the CFSK and CFSK'e (for $Q^2 > 2$ GeV$^2$) models. The calculation for $Q^2 =  3.5$ GeV$^2$ is also shown for a larger range in $x$ in the sub-view of the plot.}
\label{fig:PumplinRatio}
\end{center}
\end{figure}
Finally, as an additional cross-check of the region of validity of the CFSK'e model presented in detail above we study the condition that $\sigma_{diff} \big/ \sigma_{tot} \leq 1\big/ 2$ related to the conservation of unitarity, also called the Pumplin bound \cite{Pumplin:1973cw}. We define
\beq
R_{P} = \frac{\int d \xP \, B\left( \xP \right) F_{2 \D}^{(3)} (\beta, \xP,Q^2) }{ F_2 (x,Q^2)} \;.
\eeq
In Fig.~\ref{fig:PumplinRatio} we plot $R_P$ for several $ 0.5 \mbox{ GeV}^2 \leq Q^2 \leq 20$ GeV$^2$. We see that $R_P$ is below 0.5 down to $x \sim 10^{-12}$, which proves the validity of the model for large $Q^2$ in the kinematical region we are discussing in this paper, relevant for present and future experiments and even cosmic-ray physics.

\section{Conclusions \label{sec:Conclusions}}

We have extended the original CFSK model \cite{Capella:2000pe,Capella:2000hq} of DIS to the high-$Q^2$ region by the inclusion of QCD scaling violations via the DGLAP evolution equations. The extended model (CFSK'e) was used to calculate $F_2$, $F_L$ and $\xP \FD (\xP, \beta, Q^2)$ of the proton for $0 < Q^2 < 1600$ GeV$^2$ and all values of $x$ down to 10$^{-8}$. The agreement with existing experimental data on all of these observables is good.

The CFSK'e model for scattering off nucleons can be used for robust predictions at high energies for a wide range of observables, from DIS observables to multiplicities in heavy-ion collisions \cite{Capella:1999kv}. It also provides a bridge between perturbative and non-perturbative regimes of QCD and can serve as a baseline for deviations from the standard $Q^2$-evolution involving resummations of logarithms of $x$. Finally, it bridges unitarity corrections with the strength of diffraction at high energies.

We observed differences between our approach and high-energy QCD evolution encoded in the solution of the BK equation in kinematical regions beyond current experimental reach. At low $x$, $F_2$ at high $Q^2$ is significantly larger in our calculations. This may imply the breakdown of the collinear factorization leading to the DGLAP evolution and a transition to a linear resummation or non-linear regime. On the other hand, the BK calculation is performed at a fixed impact parameter while the CFSK explicitly includes a growing interaction radius with energy, which make it hard to quantify these deviations. At low $Q^2$, $F_L$ may shed light on the transition between the perturbative and non-perturbative (Regge) regimes.

As we have pointed out above, the model for low-$x$ structure functions at high $Q^2$ respects unitarity at the initial scale of QCD evolution, but lacks nevertheless logarithms of $1/x$ that should be resummed at high energies. One should also keep in mind that the QCD analysis presented above was done at leading order in the coupling constant. In the future we will attempt to include higher-order corrections in our calculations. An extension to the nuclear case is also under developement \cite{Armesto:2009qq}.  

\section*{Acknowledgements}
K.T. would like to thank C.~Pajares, T.~Lappi and R.~Venugopalan for illuminating conversations and P.~Newman for discussing data on diffraction. The work of N.A., C.A.S. and K.T. has been supported by Ministerio de Ciencia e Innovaci\'on of Spain under project FPA2008-01177 and contracts Ram\'on y Cajal (NA and CAS), by Xunta de Galicia (Conseller\'{\i}a de Educaci\'on) and through grant PGIDIT07PXIB206\-126PR (NA and KT), grant INCITE08PXIB296116PR (CAS) and European Commission grant PERG02-GA-2007-224770, and by the Spanish Consolider-Ingenio 2010 Programme CPAN (CSD2007-00042) (NA and CAS).

\appendix
\section{The diffractive cross section in the CFSK model}
\label{app:CFSKdiffractive}

The diffractive cross section of a virtual photon in the CFSK model \cite{Capella:2000hq} is given by three terms
\beq
\sigma^{(diff)}_{\vp p} = \sum_{i=L,S} \sigma^{(diff)}_i \,+\, \sigma_{\Pom \Pom \Pom} \;,
\eeq
where
\beq
\sigma_L^{(diff)} &=& 4 g_L^2(Q^2) \int d^2 b \, \left[ \sigma^{(tot)}_L (b,s,Q^2)\right]^2 \;, \\
\sigma_S^{(diff) } &=& 4 \sum_{T,L} \int d^2b \int_0^{r_0} d^2r \int_0^1 dz \, \left| \psi^{T,L} (z,r) \right|^2 \left[ \sigma_S (r,b,s,Q^2) \right]^2 \;, \\
\sigma_{\Pom \Pom \Pom} &=& 2 g_L^2(Q^2) \int d^2 b\, \chi_{\Pom \Pom \Pom}^L (b,s,Q^2) e^{- 2 C \chi_L (b.s.Q^2)} \nonumber \\
& & + 2 \sum_{T,L} \int d^2b \int_0^{r_0} d^2r \int_0^1 dz \, \left| \psi^{T,L} (z,r) \right|^2 \chi_{\Pom \Pom \Pom}^S (b,s,Q^2) e^{- 2 C \chi_L (b.s.Q^2)}  \;.
\eeq
Here 
\beq
\chi_{\Pom\Pom\Pom}^i(b,s,Q^2) &=& a \, \chi_i^{\Pom}(b,s,Q^2) \chi_3 (b,s,Q^2) \;,
\eeq
where $i=L,S$, $\chi_L^\Pom$ is given by the first term in eq.~(\ref{eq:TotalChi}) and $\chi_3$ is defined in eq.~(\ref{eq:TriplePomeron}). To find the (reduced) diffractive structure function as a function of $\beta$, defined by
\beq
\xP F_{2 \D}^{(3)}(\xP,\beta,Q^2) = \frac{Q^2}{4 \pi^2 \alpha_{em}} \, \int d t \, \xP \frac{d \sigma}{d \xP dt} \;,
\eeq
we perform the same decomposition as above, giving
\beq
F_{2 \D}^{(3)} = F_{2 \D \, S}^{(3)} + F_{2 \D \, L}^{(3)} + F_{2 \D \, 3\Pom}^{(3)} \;,
\eeq
where the individual pieces are extended to the whole $\beta$-region. In particular, the $S$-component amounts to
\beq
\xP F_{2\D \, S}^{(3)} = \frac{Q^2}{4 \pi^2 \alpha_{em}} \left( \sigma_S^{(0) L} \mathcal{N}\left[\tilde{\beta}^3 (1-2\beta)^2\right] + \sigma_S^{(0) T} \mathcal{N}\left[\tilde{\beta}^3(1-\beta)\right] \right) \;,
\eeq
where $\tilde \beta = (Q^2 + s_0)/(Q^2 + M^2) = \beta \tilde x / x$, and 
\beq
\mathcal{N}\left[ f(\beta) \right] = f(\beta) \Big/ \int^{\beta_{max}}_{\beta_{min}} \frac{d \beta}{\beta} f(\beta) \;,
\eeq
with $\beta_{min} = 10\, x$ and $\beta_{max} = Q^2 /(Q^2 + 4m_\pi^2)$. Next, the $L$-component is
\beq
\xP F_{2\D \, L}^{(3)} = \frac{Q^2 g_L^2(Q^2)}{4 \pi^2 \alpha_{em}} \frac{\sigma_L^{(0)}}{\left. \sigma_L^{(0)} \right|_{C=0}} \int d^2 b \, \left( \chi_L^\Pom \right)^2 \, \mathcal{N}\left[ \tilde{\beta}^{\Delta_i + \Delta_k - \Delta_\Reg} (1-\beta)^{n(Q^2)} \right] \;,
\eeq
where $\chi_L^\Pom$ is given by the first term in eq.~(\ref{eq:TotalChi}). Note the difference between our formula and \cite{Capella:2000hq} where also double $\Pom \Reg$ and $\Reg\Reg$ exchanges were taken into account (numerically, these contributions are insignificant). Finally, the $3\Pom$-component is
\beq
\label{eq:Diff3P}
\xP F_{2\D \, 3\Pom}^{(3)} &=& \xP F_{2\D \, 3\Pom}^{(3) B} \frac{\sigma_{\Pom\Pom\Pom}^{(0)}}{\left. \sigma_{\Pom\Pom\Pom}^{(0)} \right|_{C=0}} \;, \\
\xP F_{2\D \, 3\Pom}^{(3) B} &=& \frac{Q^2}{4 \pi^2 \alpha_{em}} 2a \int d^2 b \, \chi_3(b,s,Q^2,\beta)  \left[ \sigma_S^{(tot)} (s,b,Q^2) + \sigma_L^{(tot), singlet} (s,b,Q^2) \right]_{C=0} \; .
\eeq

\end{document}